\documentclass[english]{article}
\usepackage[T1]{fontenc}
\usepackage[latin9]{inputenc}
\usepackage{array}
\usepackage{textcomp}
\usepackage{mathtools}
\usepackage{multirow}
\usepackage{amsmath}
\usepackage{amsthm}
\usepackage{amssymb}
\usepackage{graphicx}
\usepackage{esint}

\makeatletter

\providecommand{\tabularnewline}{\\}

\theoremstyle{plain}
\newtheorem*{question*}{\protect\questionname}
\theoremstyle{plain}
\newtheorem{thm}{\protect\theoremname}
\theoremstyle{definition}
\newtheorem{defn}[thm]{\protect\definitionname}
\theoremstyle{plain}
\newtheorem{prop}[thm]{\protect\propositionname}
\theoremstyle{plain}
\newtheorem{cor}[thm]{\protect\corollaryname}
\theoremstyle{remark}
\newtheorem{rem}[thm]{\protect\remarkname}
\theoremstyle{plain}
\newtheorem{question}[thm]{\protect\questionname}
\theoremstyle{plain}
\newtheorem{assumption}[thm]{\protect\assumptionname}

\makeatother

\usepackage{babel}
\providecommand{\assumptionname}{Assumption}
\providecommand{\corollaryname}{Corollary}
\providecommand{\definitionname}{Definition}
\providecommand{\propositionname}{Proposition}
\providecommand{\questionname}{Question}
\providecommand{\remarkname}{Remark}
\providecommand{\theoremname}{Theorem}

\begin{document}
\title{For the use of exterior form in daily physics, an introduction without
coordinate frame.}
\author{Raphael Ducatez}

\maketitle
This is a short introduction of the exterior form formalism focus
on its applications in physics and then mostly aimed to physics students.
If exterior forms are more than a century old they are unfortunately
still seen (and teached) as a high level mathematics object and then
little used outside theorical physics. We then focus here on simple
examples which occure in daily phiysics. There exists already a lot
of very good mathematical textbooks and courses on the subject but
the originality of these notes, the physical applications aside, is
that we keep a completely geometrical approach. As a rule of a game
played here we never use a coordinate frame neither in the definitions
nor in the proofs but only at the end in order to recover the classical
physics equations. This approach is unusual but we think is helpful
for the understanding and very valuable to grab the physical meanings
of the mathematical object. To say differently we will always prefer
the following
\begin{align*}
\text{Geometric definition } & >\text{ Coordinate definition }
\end{align*}
\[
\text{Geometric proof }>\text{ Computational proof.}
\]

A large part of these notes are just ``notations rewriting'' of
well known physical objects but this should not be underestimate as
it gives short and elegant expressions that are useful both for insights
and computations. Appart from the \emph{game} explained above most
the material presented here is very well known \cite{frankel2011geometry,flanders1963differential,burke1985applied}
with exeption maybe of Corollary \ref{cor:conserve-1f} for which
we have surprisingly not found its statement in the litterature and
the discussion around Corollaries \ref{cor:associated_field}, \ref{cor:wave_equation}
which we believe deserves more publicity. Hopefully the approach presented
here could be generalized in a very abstract setting. 

\section{Exterior forms}

\subsection{Submanifolds as elementary objects}

We denote $\Omega$ as the ``physical universe'' and we will always
think $\Omega$ as a 3 or 4 dimensional manifold ($\mathbb{R}^{3}$
or $\mathbb{R}^{4}$ for example). Our basic objects will be submanifolds
of $\Omega$
\[
{\cal V}\subset\Omega
\]
In any textbook the next step is to attach to $\Omega$ a set of local
coordinate. This is the step we will try to avoid here, or at least
we will try to write down as much as we can without any choice of
local coordinate. For example if one consider the church ``Notre
Dame'', one could say that it is at 48° North-2° East on the surface
of the Earth but one could also say that the church is just a volume
embedded in the universe ${\cal V}_{\text{Notre Dame}}\subset\Omega$
and this is a more fundamental mathematical representation. In these
notes we ask the following.
\begin{question*}
If we restrict ourself to work only with submanifold without local
coordinate, what could we do from a mathematical point of view? 
\end{question*}
Here we will use exterior forms but will try to avoid Grassmann algebra
as well.
\begin{center}
\begin{figure}
\centering{}\includegraphics[width=8cm]{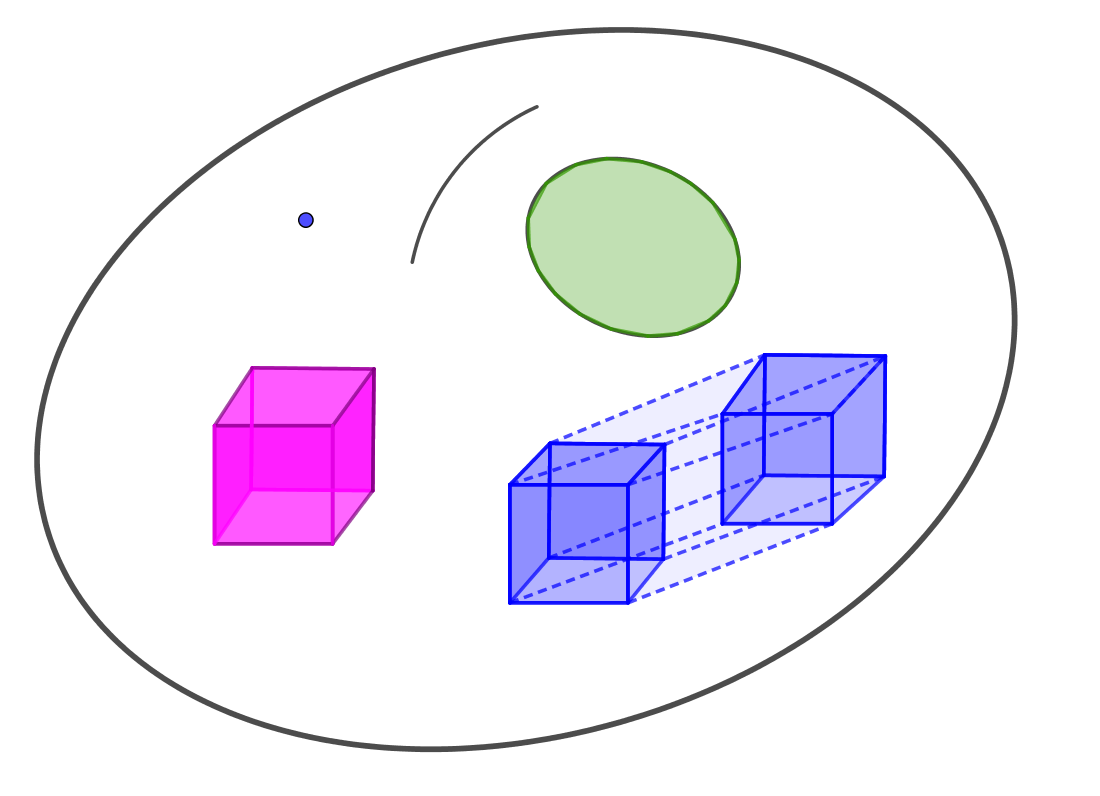}\caption{\label{manifold} What can we do if we just have submanifolds and
no coordinate frame ?}
\end{figure}
\par\end{center}

\subsection{Integration on a submanifold}

Usually, introduction about exterior form start at the coordinate
level giving the formal definition of an exterior algebra \cite{paulin2007geometrie,CodbillonElementTopologie,lee2009manifolds}.
However, it is not so obvious at first sight why such an object is
usefull and interesting in physics. So we would instead propose the
following definition to go immediatly to the main point \cite{flanders1963differential}.
\begin{defn}
A $k-$form is \emph{what }to be integrated on a $k-$dimension submanifold.
\label{def:exterior-form}
\end{defn}

This definition is made vague and unformal on purpose. Because it
much closer of what any physics would want as a mathematical object.
That is to choose an submanifold\footnote{More precisely an ``oriented submanifold''. The integration on path
from A to B or from B to A have a different sign. Similarly the integration
of a flow through a surface going inward or outward have a different
sign.} : a path, a surface, a volume or a time interval $\times$ a volume
and have physical quantity to integrate on it. This should really
be though as the ``meaningful physical quantity'', as it doesn't
depend on the choice of coordinate, not even the space-time metric
$g$. If one think of a glass of water the quantity of water in the
volume defined by the glass (a 3 dimension submanifold) should exist
and be independant on the methods we measure space or time. From the
lowest mathematical level this is just an application $\alpha$ that
for any $k-$dimension submanifold ${\cal V}$ associate a real number
$\alpha({\cal V})\in\mathbb{R}$.
\begin{align*}
\alpha:k\text{\text{-dimension submanifold}} & \rightarrow\quad\mathbb{R}\\
\qquad\quad{\cal V}\qquad\quad & \rightarrow\quad\alpha({\cal V})
\end{align*}
 We also denote
\[
\alpha({\cal V})=\int_{{\cal V}}\alpha
\]
but this notation is best to be seen as the definiton of the r-h-s
than of the l-h-s.

Here are some examples in the case $\Omega$ is 3 dimensional.
\begin{center}
\begin{tabular}{|c|c|c|c|}
\hline 
dimension & basis & quantity & unit\tabularnewline
\hline 
\hline 
\multirow{4}{*}{0-form} & \multirow{4}{*}{$1$} & Temperature & $K$\tabularnewline
\cline{3-4} \cline{4-4} 
 &  & Pressure & $Pa$\tabularnewline
\cline{3-4} \cline{4-4} 
 &  & Potential & $V$\tabularnewline
\cline{3-4} \cline{4-4} 
 &  & Reflective index & $1$\tabularnewline
\hline 
\multirow{5}{*}{1-form} & \multirow{5}{*}{$dx,dy,dz$} & Temperature gradient & $K.\boldsymbol{m^{-1}}$\tabularnewline
\cline{3-4} \cline{4-4} 
 &  & Pressure gradient & $Pa.\boldsymbol{m^{-1}}$\tabularnewline
\cline{3-4} \cline{4-4} 
 &  & Electric field & $V.\boldsymbol{m^{-1}}$\tabularnewline
\cline{3-4} \cline{4-4} 
 &  & Vector potential & $Vs.\boldsymbol{m^{-1}}$\tabularnewline
\cline{3-4} \cline{4-4} 
 &  & Magnetizing field & $Cs^{-1}.\boldsymbol{m^{-1}}$\tabularnewline
\hline 
\multirow{5}{*}{2-form} &  & mass flow & $kg.s^{-1}.\boldsymbol{m}^{-2}$\tabularnewline
\cline{3-4} \cline{4-4} 
 & $dx\wedge dy$,  & charge flow & $C.s^{-1}.\boldsymbol{m}^{-2}$\tabularnewline
\cline{3-4} \cline{4-4} 
 & $dx\wedge dz$,  & radiation & $W.\boldsymbol{m}^{-2}$\tabularnewline
\cline{3-4} \cline{4-4} 
 & $dy\wedge dz$ & magnetic field & $Vs.\boldsymbol{m^{-2}}$\tabularnewline
\cline{3-4} \cline{4-4} 
 &  & Electric displacement field & $C.\boldsymbol{m}^{-2}$\tabularnewline
\hline 
\multirow{4}{*}{3-form} & \multirow{4}{*}{$dx\wedge dy\wedge dz$} & density of mass & $kg.\boldsymbol{m}^{-3}$\tabularnewline
\cline{3-4} \cline{4-4} 
 &  & density of charge & $C.\boldsymbol{m}^{-3}$\tabularnewline
\cline{3-4} \cline{4-4} 
 &  & density of energy & $J.\boldsymbol{m}^{-3}$\tabularnewline
\cline{3-4} \cline{4-4} 
 &  & heat capacity & $JK^{-1}.\boldsymbol{m}^{-3}$\tabularnewline
\hline 
\end{tabular}
\par\end{center}

In $\mathbb{R}^{3}$ the dimension of the $k-$form $\Lambda^{k}(\mathbb{R}^{3})$
are respectively $1,3,3$ and $1$. The main message here is that
exterior form are natural to describe daily physics quantities and
are not more complicated to use than the standard scalars of vector
fields. We still stress a few differences. What are usual called ``scalars''
are here distinguished between 0-form and 3-form in a very similar
way as we distinguish intensive and extensive properties. And what
are usual called ``vector fields'' are distinguished between 1-form
and 2-form. It is still interesting to make such a difference at the
mathematical level since they would not behave the same way with a
change of coordinate. A fact that stills appears in the system of
units.

In $\mathbb{R}^{4}$ the dimension of the $k-$form $\Lambda^{k}(\mathbb{R}^{3})$
are respectively $1,4,6,4,$ and $1$. The $0$-forms and $4$-forms
are the usual scalars while $1$-forms and $3$-forms are the usual
4-vectors. The 2-form are less commun but correspond to anti-symmetric
2-tensors. Generally speaking anti-symmetric properties of tensor
are related to integration so have some geometric meaning. Here are
some examples of physical quantities written as exterior forms in
4-dimensional.
\begin{center}
\begin{tabular}{|c|c|c|c|}
\hline 
dimension & basis & quantity & unit\tabularnewline
\hline 
\multirow{1}{*}{0-fom} & \multirow{1}{*}{$1$} & (same as in $\mathbb{R}^{3}$) & $*$\tabularnewline
\hline 
\multirow{2}{*}{1-form} & $dx,dy,dz,$ & 4-gradient  & $*.\boldsymbol{m^{-1}}$ or \tabularnewline
 & $dt$ & $(\partial_{t}f,\nabla f)$ & $*.\boldsymbol{s^{-1}}$\tabularnewline
\hline 
\multirow{2}{*}{2-form} & $dx\wedge dy,\cdots$ & Electromagnetic Field & $*.\boldsymbol{m^{-2}}$ or\tabularnewline
 & $dt\wedge dx,\cdots$ & $F=(E,B)$  & $*.\boldsymbol{m^{-1}s^{-1}}$ \tabularnewline
\hline 
\multirow{2}{*}{3-form} & $dx\wedge dy\wedge dz$,... & Density and flow  & $*.\boldsymbol{m}^{-3}$ or\tabularnewline
 & $dt\wedge dx\wedge dy$,... & $J=(\rho,j)$ & $*.\boldsymbol{m}^{-2}\boldsymbol{s}^{-1}$\tabularnewline
\hline 
4-form & $dt\wedge dx\wedge dy\wedge dz$ & Field Lagrangian & $*.\boldsymbol{m}^{-3}\boldsymbol{s}^{-1}$\tabularnewline
\hline 
\end{tabular}
\par\end{center}

For example integrating $J=j\,dt\wedge dx\wedge dy$ would give the
amount of matter that has been through a (horizontal) surface within
a time interval.

\subsection{Tangent vector fields}

``Vector field'' is also usually used to design another mathematical
object that to avoid ambiguity we will call here \emph{tangent vector
fields.} If exterior forms are associated with integration, one intead
should think of tangent vector fields \cite{lee2009manifolds} informally
as \emph{what} describes a flow, ie a transport. 
\begin{defn}
A flow is a familly of application on $\Omega$ 
\[
\phi_{t}:\Omega\rightarrow\Omega\quad t\in\mathbb{R}_{+}.
\]
\end{defn}

We should stress here that $\phi_{t}$ is well defined at a geometrical
level. And even if it looks abstract, it is very simple and natural
because it is nothing but the description of how the system has evolved
at different times. Moreover if we assume that is a semi-group 
\[
\phi_{t_{1}+t_{2}}=\phi_{t_{1}}\circ\phi_{t_{2}}\quad t_{1},t_{2}\in\mathbb{R}_{+}
\]
one can introduce a tangent vector field such the flow is the solution
of a differential equation.
\begin{defn}
For a flow $\phi_{t}$ the associated tangent vector field $X$ is
such that for any for point $x_{0}\in\Omega$, 
\begin{equation}
x(t)=\phi_{t}(x_{0})\quad\Leftrightarrow\quad\begin{cases}
x(0)=x_{0}\\
\partial_{t}x(t)=X(x(t)).
\end{cases}\label{eq:Flow}
\end{equation}
\end{defn}

One could see the tangent vector field as the derivative of the flow
$"X=\frac{d}{dt}\phi_{t}"$. So Equation (\ref{eq:Flow}) can be thought
as the usual way to measure a velocity flow which is to observe the
system at different time and ``differentiate'' this evolution. The
main message here is that a tangent vector field it is a completely
different mathematical object than the exterior forms. It is used
to describe different physical notions and has different basis notation
and system of units as for example :
\begin{center}
\begin{tabular}{|c|c|c|c|}
\hline 
object & basis & quantity & unit\tabularnewline
\hline 
\hline 
tangent vector & $\partial_{x},\partial_{y},\partial_{z}$ & velocity of a fluid & $\boldsymbol{m}.s^{-1}$\tabularnewline
\hline 
\end{tabular}
\par\end{center}

As illustrated by the system of units tangent vector fields are usually
refered as covariant tensors while exterior form are contravariant
tensors. In $\mathbb{R}^{4}$, we also consider transport in the time
direction and use the basis element $\partial_{t}$. For example one
can consider a motionless object but still think of it as transported
from the past toward the future. In general we see the evolution from
a point $(t_{0},x_{0},y_{0},z_{0})$ to another point $(t_{1},x_{1},y_{1},z_{1})$
as a transport both in time and space. 

\subsection{Transported and enlarged submanifold}

Tangent vector fields allow us to define a few opérations on a submanifold
${\cal V}\subset\Omega$ which are, we recall, our basic object of
interest.
\begin{defn}
With $X$ a tangent vector field associated to a semi-group $\phi_{t}$
and ${\cal V}$ a submanifold, we define
\begin{itemize}
\item An transported submanifold 
\[
{\cal V}(t)=\phi_{t}({\cal V})=\{x(t),\,x_{0}\in{\cal V}\}
\]
with $x(t)$ solution of (\ref{eq:Flow}).
\item An enlarged submanifold 
\[
{\cal I}_{t}({\cal V})=\bigcup_{s\in[0,t]}\phi_{s}({\cal V})=\{x(s),\,x_{0}\in{\cal V},\,s\in[0,t]\}
\]
that is a submanifold which is one dimensional larger than ${\cal V}$.
\end{itemize}
\end{defn}

For example if one think of $X$ describing the velocity field of
flow of water, then a transported submanifold would be used to describe
anything that is dived in the water and then carried away by the flow.
For the enlarged submanifold one can see it as keeping track of the
successive positions of transported manifold along the flow as in
an numerical simulation. For example in the case of a point ${\cal V}=\{x_{0}\}$,
this is just the line traced by the trajectory $x(t)$. In the case
of the leave it is the volume in the water that has been covered by
the leave on its way. 

Both ${\cal V}(t)$ and ${\cal I}_{t}({\cal V})$ are natural objects
on which integrate an exterior form. This is will be presented in
the next section.
\begin{center}
\begin{figure}
\centering{}\includegraphics[width=8cm]{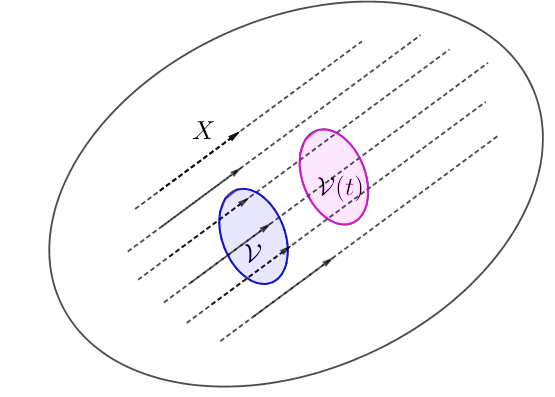}\caption{\label{transported-submanifold} The transported submanifold ${\cal V}(t)$.}
\end{figure}
\par\end{center}

.

\begin{figure}
\centering{}\includegraphics[width=8cm]{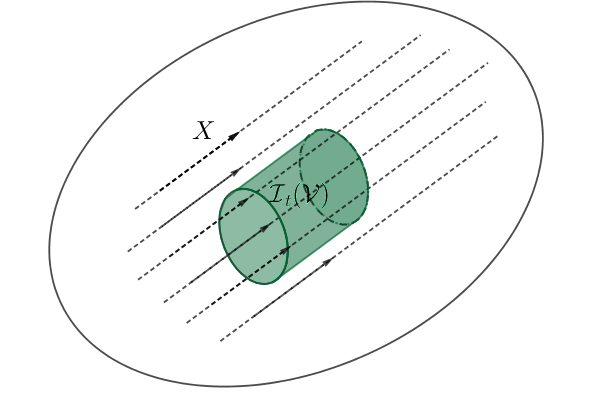}\caption{\label{enlarged-submanifold} The enlarged submanifold ${\cal V}(t)$.}
\end{figure}

\subsection{Pullback transport, Lie derivative and interior product}

We now present some basic operations to manipulate exterior form as
appears in math textbooks \cite{jost2008riemannian,lee2009manifolds,frankel2011geometry}.
All these operations also have a simple definition in a coordinate
system but the goal of these notes is to stay at a geometric point
of view and to give an motivation to introduce these objects. Also
it makes it clear that they are independent of the choice of the coordinate
and therefore should have some physical meaning.
\begin{defn}
With $X$ a tangent vector field associated to a semi-group $\phi_{t}$
and $\alpha$ a $k$ form we introduce the following quantities
\begin{itemize}
\item The \emph{pullback transport} $\phi_{t}^{*}(\alpha)$ defined as a
$k$ form such that for any $k$ dimensional submanifold ${\cal V}$
\[
[\phi_{t}^{*}(\alpha)]({\cal V})=\alpha({\cal V}(t)).
\]
\item The \emph{Lie derivative} $L_{X}\alpha$ defined as\emph{ }
\[
L_{X}\alpha=\frac{d}{dt}|_{t=0}\phi_{t}^{*}(\alpha)
\]
\end{itemize}
\end{defn}

Remark that the Lie derivative is then defined as the $k$ form such
that for any $k$ dimensional submanifold ${\cal V}$
\[
[L_{X}\alpha]({\cal V})=\lim_{t\rightarrow0}\frac{\alpha({\cal V}(t))-\alpha({\cal V}(0))}{t}.
\]

\begin{defn}
With $X$ a tangent vector field associated to a semi-group $\phi_{t}$
and $\alpha$ a $k$ form we introduce the following quantities
\begin{itemize}
\item A $k-1$ form $I_{t}\alpha$ defined such that for any $k-1$ dimensional
submanifold ${\cal V}$
\[
[I_{t}\alpha]({\cal V})=\alpha({\cal I}_{t}({\cal V}))
\]
with ${\cal I}_{t}({\cal V})$ the enlarged submanifold.
\item The \emph{interior product} \emph{$i_{X}\alpha$ defined as }
\[
i_{X}\alpha=\frac{d}{dt}|_{t=0}I_{t}\alpha
\]
\end{itemize}
\end{defn}

That is the $k-1$ form such that for any $k-1$ dimensional submanifold
${\cal V}$
\[
[i_{X}\alpha]({\cal V})=\lim_{t\rightarrow0}\frac{\alpha({\cal I}_{t}({\cal V}))}{t}.
\]

This has a natural physical meaning : for example with $\rho$ a density
of matter (a 3-form) and $X$ a velocity field, $i_{X}\rho$ (a 2-form)
describes the density flow (For any surface $S$, $i_{X}\rho(S)$
is the flow of matter that is going through $S$).

The approach presented here is actually very general : any operation
on submanifolds can be translated into an operation on forms. Indeed
considering an application that from a $k$ submanifold ${\cal V}$
gives another submanifold ${\cal V}\rightarrow\tilde{{\cal V}}$.
We can define an application that from a $k$ form $\alpha$ gives
another form $\alpha\rightarrow\tilde{\alpha}$ such that

\begin{equation}
\tilde{\alpha}(\tilde{{\cal V}})=\alpha({\cal V})\label{eq:DualForm-1}
\end{equation}
Equation (\ref{eq:DualForm-1}) can also be used to have ``dual''
application $\tilde{\alpha}\rightarrow\alpha$ that from a form $\tilde{\alpha}$
define a $k$ form $\alpha$. 
\begin{question*}
If we restrict ourself to the use of this operators, what physics
equations could we write ?
\end{question*}
We will work on that question later but first we have to introduce
the exterior derivative.

\section{Exterior Derivative}

\label{sec:Exterior-Derivative}

\subsection{Integration on the boundary}

The usual formal definition of the exterior derivative use the derivation
at the coordinate level \cite{jost2008riemannian,lee2009manifolds,frankel2011geometry}
but as before it is not very clear at first sight why such an objet
would be interesting in physics. So here again is another definition
that is informal but that looks very natural and capture the main
interest of the object.
\begin{defn}
\label{def:ExteriorForm}For a $k-$form $\alpha$, the exterior derivative
$d\alpha$ is defined as the $(k+1)-$form such that for any $(k+1)$
dimensional submanifold ${\cal V}$
\[
d\alpha({\cal V})=\alpha(\partial{\cal V}).
\]
\end{defn}

\begin{center}
\begin{figure}
\centering{}\includegraphics[width=8cm]{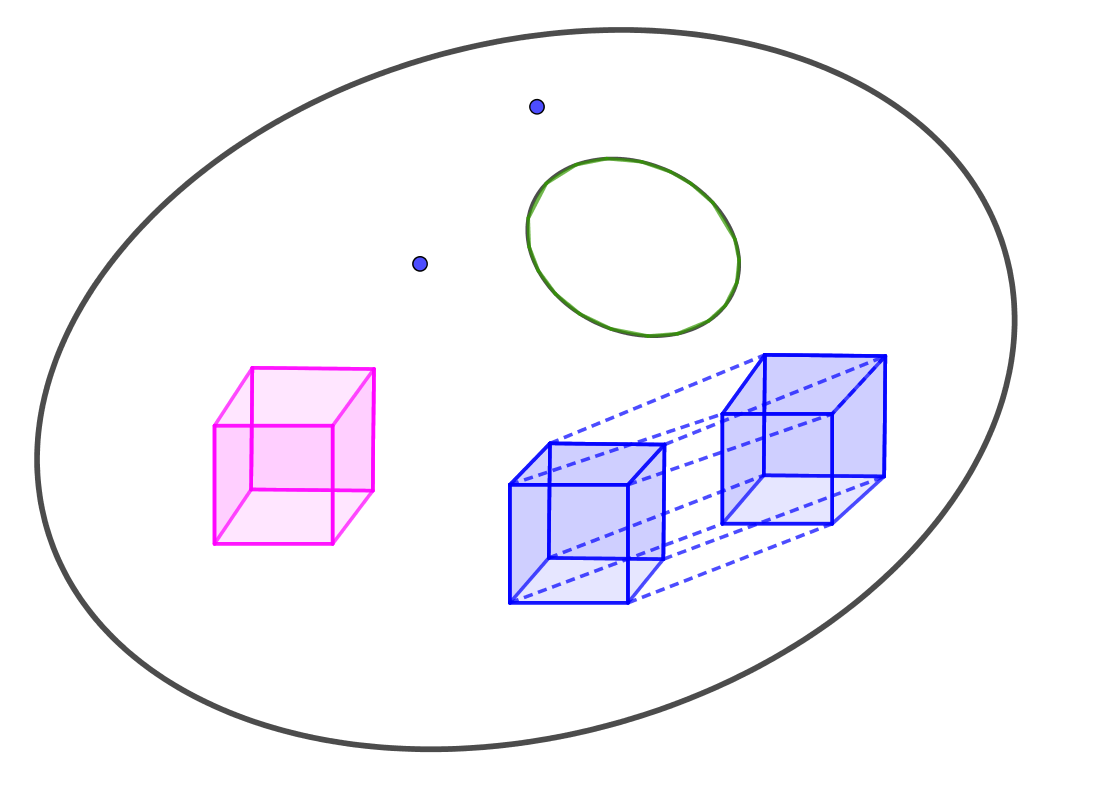}\caption{\label{Boundaries} Boundaries of the submanifolds in Figure \ref{manifold}.}
\end{figure}
\par\end{center}

This is of course the very famous Stokes Theorem. 
\[
\int_{{\cal V}}d\alpha=\int_{\partial{\cal V}}\alpha
\]
but as in Definition \ref{def:exterior-form} such an equation could
also be seen as a definition of an application on the $(k+1)-$submanifolds,
that is a $(k+1)-$form and such a definition does not depend on a
choice coordinate. 

Also boundaries of submanifold are very commun objects : the extremal
points of a path, the circle surounding a disc, the surface around
a volume. One can also think of the initial and final configurations
of a system as the $3$ dimension boundaries of the evolving system
in time and space seen as a 4 dimension submanifold. 

In a $3$ dimension space whether the form $\alpha$ is a $0,1$ or
$2$-form, the exterior derivative $d\alpha$ is called gradient,
curl and divergence\footnote{The most commun convention use the metric tensor to transform these
forms into vector field. See Section \ref{subsec:star-operator}.}. 

\[
\Lambda^{0}(\mathbb{R}^{3})\xrightarrow{\text{grad}}\Lambda^{1}(\mathbb{R}^{3})\xrightarrow{\text{curl}}\Lambda^{2}(\mathbb{R}^{3})\xrightarrow{\text{div}}\Lambda^{3}(\mathbb{R}^{3}).
\]
where $\Lambda^{k}(\mathbb{R}^{3})$ denote the space of $k-$form.

\subsection{Conserved quantity}

With the previous section in $\mathbb{R}^{3}$ a divergence-free vector
field is then a $2$-form $\alpha$ such that $d\alpha=0$. In $\mathbb{R}^{4}$
we can propose a similar definition of a conserved quantity. If we
go back to our example of evolving system in time and space. For conservation
we ask that the integration on the initial and final configurations
gives the same result. This is guaranty if $d\alpha=0$ so we can
propose the following definition.
\begin{defn}
\label{def:conserved-quantity}A conserved quantity is a $3-$form\footnote{And more generaly a $(n-1)-$form if $\Omega$ is $n$ dimensional}
$J$ such that $dJ=0$.
\end{defn}

Writting the 3-form in a coordinate frame $J=(\rho,j_{x},j_{y},j_{z})$,
the condition $dJ=0$ is just the Continuous Equation \cite{flanders1963differential}
\[
\partial_{t}\rho+\text{div}(j)=0.
\]
An advantage of the above formulation is that it does not depends
on the coordinate frame or the metric. Equivalently for any ${\cal V}$
we have $J(\partial{\cal V})=0$. In word : ``What is going into
${\cal V}$ is the same as what is going out of ${\cal V}$'' which
is also a very natural non mathematical definition of what a conserved
quantity is.

\subsection{Gauge Invariance}

We also state the following important observation : «the boundary
of a manifold has no boundary» $\partial(\partial{\cal V})=\emptyset$.
Therefore for any form $\alpha$, we have 
\[
(d\circ d\alpha)({\cal V})=\alpha(\partial(\partial{\cal V}))=0
\]
 and we write the following proposition.
\begin{prop}
$d\circ d=0$
\end{prop}

For example in $\mathbb{R}^{3}$ this is the well known $\text{curl}\circ\text{grad}=0$
and $\text{div}\circ\text{curl}=0$. 
\begin{cor}
(Gauge invariance) For a $k$-form $\alpha$ and any $(k-1)$ form
$\beta$ we have
\[
d(\alpha+d\beta)=d\alpha.
\]
\end{cor}

If the physical quantity of interest is $d\alpha$, then the $k$-form
$\alpha$ is not a unique and any $\alpha+d\beta$ works as well.
Setting a particular $\beta$ is to make a Gauge choice. For example
for the magnetic field $B=\text{curl}(A+\text{grad}(f))$.

\subsection{Cartan's magic formula}

We go back to the math textbooks \cite{jost2008riemannian,lee2009manifolds,frankel2011geometry}.
Here are some remarks related to the boundaries (see for example Figure
\ref{transported-submanifold}) : The boundary of the transported
submanifold is just the transport of the boundary of the inital submanifold:
\[
\partial[{\cal V}(t)]=[\partial{\cal V}](t).
\]
As a consequence we have this Proposition. 
\begin{prop}
With $X$ a tangent vector field associated to a flow $\phi_{t}$
and $\alpha$ a $k$ form we have
\[
\phi_{t}^{*}(d\alpha)=d\phi_{t}^{*}(\alpha)\quad\text{and}\quad dL_{X}\alpha=L_{X}(d\alpha).
\]
\end{prop}

We also remark following. The boundary of ${\cal I}_{t}({\cal V})$
has three terms : ${\cal V}$, ${\cal V}(t)$ and the union of $\partial{\cal V}(s)$
for $s\in[0,t]$. The latter also corresponds to ${\cal I}_{t}(\partial{\cal V})$.
Therefore we have\footnote{The orientation of the boundary parts is a bit tricky.}
\[
[I_{t}d\beta]({\cal V})=d\beta({\cal I}_{t}({\cal V}))=\beta({\cal V}(t))-\beta({\cal V}(0))-\beta({\cal I}_{t}(\partial{\cal V}))
\]
 and then obtain the relation known as Cartan's magic formula.
\begin{prop}
\label{prop:(Cartan's-magic-formula)}(Cartan's magic formula) With
$X$ a tangent vector field associated to a flow $\phi_{t}$ and $\alpha$
a $k$ form we have

\begin{equation}
I_{t}d\alpha=\phi_{t}^{*}(\alpha)-\alpha-dI_{t}\alpha\quad\text{and}\quad L_{X}=d\circ i_{X}+i_{X}\circ d.\label{eq:Cartan}
\end{equation}
\end{prop}

\begin{figure}
\centering{}\includegraphics[width=4cm]{tangent_vector_3_b}\includegraphics[width=4cm]{tangent_vector_3_a}\includegraphics[width=4cm]{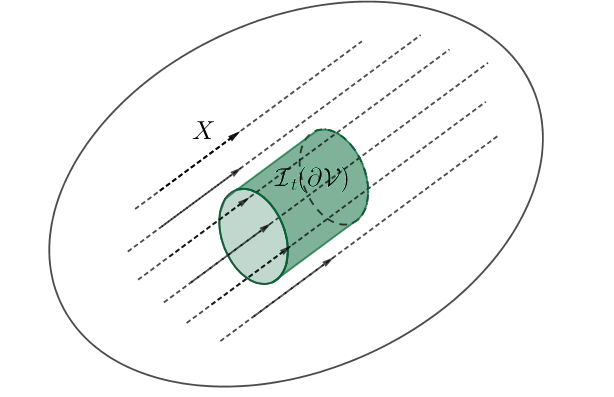}\caption{\label{enlarged-submanifold-1} A ``Geometric proof'' of Cartan's
formula $i_{X}\circ d=L_{X}-d\circ i_{X}$ seen as a consequence of
$\partial[{\cal I}_{t}({\cal V})]={\cal V}(t)\cup{\cal V}(0)\cup{\cal I}_{t}(\partial{\cal V})$
: $i_{X}\circ d$ is associated to the figure on the left, $L_{X}$
to the figure in the center and $d\circ i_{X}$ to the figure on the
right.}
\end{figure}

\subsection{Lie Derivative and Hydrodynamics}

In the following small second sections, we add two applications of
the Lie Derivative. First in the context of fluid hydrodynamics, it
is natural to consider exteriors forms $\alpha_{t}$ that evolve in
time, such that for transported submanifolds ${\cal V}(t)$, the quantity
$\alpha_{t}({\cal V}(t))$ stay constant. We would then have $\alpha_{t}=\phi_{-t}^{*}(\alpha_{0})$.
And if we add some modifications of $\alpha_{t}$ along the way, we
obtain the following equation.
\begin{defn}
(Euler Equation) $\partial_{t}\alpha+L_{X}\alpha="\text{additional terms"}$\footnote{Such as exterior forces, pressure, viscosity, source...}.
\end{defn}

Here again, our main message is that one need to be careful whether
$\alpha$ is a 0,1,2,3 form to compute $L_{X}\alpha$ in a coordinate
system. W can write
\[
L_{X}\alpha=\sum_{i=1}^{3}X_{i}L_{e_{i}}\alpha+\sum_{i=1}^{3}dX_{i}\wedge i_{e_{i}}\alpha.
\]
 with $X=\sum_{i=1}^{3}X_{i}e_{i}$ and the first term is the usual
$\vec{X}.\vec{\text{grad}}$ that appears in the Lagrangian derivative.
We could call it a ``translation term''. For a general tangent vector
field $X$, the submanifolds are deformed along the flow (rotation,
dilatation,...) and we need to add what we could call a ``deformation
term'' for which we can use the following table.
\begin{center}
\begin{tabular}{|c|c|c|}
\hline 
dimension & deformation term & examples of equations\tabularnewline
\hline 
$0$-form $f$ & $\emptyset$ & Lagrangian derivative\tabularnewline
\hline 
$1$-form $A$ & $(\nabla X)A$ & Deformation gradient\tabularnewline
\hline 
$2$-form $B$ & $(B.\nabla)X-(\nabla.X)B$ & Cauchy momentum equation, \tabularnewline
 &  & the vorticity equation\tabularnewline
\hline 
$3$-form $\rho$ & $\rho(\nabla.X)$ & Continuity equation\tabularnewline
\hline 
\end{tabular}
\par\end{center}

Where we use the notations $\nabla X=(\partial_{i}X_{j})_{1\leq i,j\leq3}$
for the Jacobian, $\nabla.X=\sum_{i=1}^{3}\partial_{i}X_{i}$ for
the divergence and $B.\nabla=\sum B_{i}\partial_{i}$ for the Lagrangian
derivative.

\subsection{Lie Derivative and forces applied at the boundary}

To consider the total force applied on a submanifold ${\cal V}$ one
would simply write $\overrightarrow{\boldsymbol{F}}=\int_{{\cal V}}\overrightarrow{F}$.
Unfortunatly if ${\cal V}\subset\Omega$ a general manifold ($\neq\mathbb{R}^{d})$
one can't just sum vectors that are defined at different points and
the last equation doesn't make any sense. To overcome this difficulty
we will consider a displacement described by a tangent vector field
$X$ and consider the work generated by the force as the displacement
is made. We call the quantity $F_{X}=\int_{{\cal V}}\overrightarrow{F}.\overrightarrow{X}$
the force in $X$-direction. For example in the case $\Omega=\mathbb{R}^{3}$
and a constant vector field $\overrightarrow{X}=\overrightarrow{e_{x}},\overrightarrow{e_{y}}$
or $\overrightarrow{e_{z}}$ (that describe a translation) one recover
the standard definition $\overrightarrow{\boldsymbol{F}}=(F_{x},F_{y},F_{z})$
where $F_{x}=\int_{{\cal V}}\overrightarrow{F}.\overrightarrow{e_{x}}$.
But here $F_{X}$ is a very general notion because $X$ can then be
any tangent vector field. For example in the case of $X$ are the
rotations it gives the moment of force. And more generally it could
be the force associated to any deformation as for example a dilatation
or a contraction. This is the quantity that can be generalized in
the exterior form formalizm.
\begin{defn}
\label{def:force}If $\alpha$ describes the energy of a system and
$X$ is a tangent vector field then $L_{X}\alpha$ is the associated
force in the $X$-direction.
\end{defn}

We then have a $0$,$1$,$2$ or $3$ form whether it is a ponctual
force, a linear force, a surface force or a volume force.

Here is the motivation for such a definition. If ${\cal V}$ is submanifold
that describe a physical object and $\alpha({\cal V})$ is the energy
of the system. The tangent vector field $X$ defined a displacememt
of the object as the transported manifold ${\cal V}(s)$. Then the
variation of the energy is given by
\[
-\frac{d}{ds}\alpha({\cal V}(s))=\frac{d}{ds}[\phi_{s}^{*}\alpha]({\cal V})=(L_{X}\alpha)({\cal V}).
\]

\begin{rem}
If $d\alpha=0$ then\footnote{Cartan formula} $(L_{X}\alpha)({\cal V})=(i_{X}\alpha)(\partial{\cal V})$.
\end{rem}

We mention here a few examples. 
\begin{itemize}
\item Archimedes principle : For the pressure which is a $3-$form $P$
that describes the energy per volume. Then the force on a volume ${\cal V}$
is $L_{X}P({\cal V})$ which may correspond to the weight of the fluid
and is also given by $(i_{X}P)(\partial{\cal V})$ which is pressure
force integrated at the surface.
\item Magnetic force : For the magnetic field $B$ is a 2-form and a surface
$S$ for which $\partial{\cal S}$ is a closed loop with a circulating
electric courant $i_{0}$, the energy is given by $i_{0}B({\cal S})$
and then the force is $i_{0}L_{X}B({\cal S})$ which is the variation
of the magnetic flux. But it is also given by $i_{0}(i_{X}B)(\partial{\cal S})$
which in a coordinate system reads $i_{0}\oint_{\partial{\cal S}}\overrightarrow{X}.\overrightarrow{B}\boldsymbol{\times}\overrightarrow{d\ell}$
(Lorentz force).
\end{itemize}

\section{The metric and Maxwell equations}

\subsection{The volume form}

Surprisingly so far we only have a manifold $\Omega$ and didn't attached
any metric on it\footnote{For mathematicians we use differential geometry but not Riemanian
geometry}. But of course it is of fundamental importance in physics. Formally
a metric symetric 2-tensor $g_{ij}$ so it is neither a form nor a
tangent vector. The first use of the metric for differential form
is the definition of a ``volume form'' $\nu$. 
\begin{defn}
A metric $g$ locally gives a orthonormal basis $dx_{1},dx_{2},\cdots,dx_{n}$
and we define
\[
\nu=dx_{1}\wedge dx_{2}\wedge\cdots\wedge dx_{n}.
\]
\end{defn}

For example once you have defined the meter and what orthogonal means
then you can construct the meter-cube. And then have a $3$-form $\nu$
such that for any 3 dimensional submanifold ${\cal V}$ 
\[
\nu({\cal V})=\text{"Volume of }{\cal V}\text{ in }m^{3}\text{"}.
\]

\subsection{The $\star$ operator}

\label{subsec:star-operator}From a purely mathematical point of view
of the exterior forms the metric also appears in a quite indirect
way. Remark that the space of $k$-form has the same dimension as
the space of $(n-k)$ form. They are therefore isomorphic. The $g$
metric gives a method to construct such an isomorphism that is called
the Hodge $\star$ operator. Again a metric $g$ locally gives a orthonormal
basis $dx_{1},dx_{2},\cdots,dx_{n}$ and to give a very simplified
definition.
\begin{defn}
That $\star$-operator associates a $k-$form to a $(n-k)$-form that
is the ``complement'' in the basis\footnote{The formula is just wrong because we do not precise the sign which
depend on the order of the $dx_{i}$ and on the signature of $g$.
But from the beginning we didn't precise the orientation of the submanifold
neither... We refer to the previously mentioned textbook for a correct
definition.} : 
\[
\star\left(\bigwedge_{i\in J}dx_{i}\right)=\pm\bigwedge_{i\notin J}dx_{i}\quad\text{for \ensuremath{J\subset\{1,\cdots,n\}} with k element }
\]
\end{defn}

For example : $\star\,(dx)=dy\wedge dz\wedge dt$ and $\star\,(dx\wedge dt)=dy\wedge dz$.
Notice that in physics it happens that we use the same name for objects
that may be different at a formal mathematical level. For example
one usually doesn't make a difference between a density $\rho=\rho(x)\nu$
that is a $3$-form, and the function $\rho(x)$ is a $0$-form. Or
a ``vector field $X$'' may refer to the tangent vector field, or
to $n-1$-form $i_{X}\nu$ induced by this tangent vector field, or
to the 1-form $\alpha=\star i_{X}\nu$.

\[
\begin{array}{ccccc}
1\text{-form} &  & (n-1)\text{-form} &  & \text{tangent vector}\\
\star\alpha & \leftrightarrow & \alpha\,=\,i_{X}\nu & \leftrightarrow & X
\end{array}
\]
In Newton second law for instance, we identify a acceleration which
is a tangent vector to a force which is a $1$ form ($ma=F=\text{grad}(f)$).
In all these transformations, to be rigorous one should add the metric
$g$ and/or the factor $\sqrt{\text{det}g}$. If $g$ is the identity
matrix, this is of course never written down, but it has to be done
for example in relativity when ``moving up and down indices'': $J^{\mu}\rightarrow J_{\mu}$
or $F^{\mu\nu}\rightarrow F_{\mu\nu}$.

\subsection{The Laplace-De Rham Operator}

We now define the Laplace-De Rham Operator.
\begin{defn}
(Laplace-De Rham) 
\[
\Delta=\partial\circ d+d\circ\partial
\]
 where $\partial=\star\circ d\circ\star$. It is an operator that
from a $k$-form gives an $k$-form.
\end{defn}

In $\mathbb{R}^{3}$ and for a $0-$form $f$ (or $3$-form), because
$\partial f=0$ this is the usual formula 
\[
\Delta f=\text{div}(\text{grad}(f))
\]
and for a $1$-form $u$ (or $2$-form), we have the well known\footnote{But seen here as the definition of $\Delta u$.}
\[
\Delta u=\text{grad}(\text{div}(u))-\text{curl}(\text{curl}(u)).
\]
 For a general manifold $\Omega$ with metric $g$ and a $0-$form
$f$, $\Delta f=\left(\star\,d\star d\right)f$ which reads 
\[
\Delta f=\frac{1}{\sqrt{\text{det}g}}\sum_{ij}\partial_{j}(\sqrt{\text{det}g}g^{ij}\partial_{i}f).
\]
In the Minkovski space $\mathbb{R}^{1,3}$, because $g$ is not positive
we obtain d'Alembert operator and denote $\square=\partial\circ d+d\circ\partial$
instead. In the case $d\alpha=0$ (or $\partial\alpha=0$) we may
also denote $\square\alpha=\star\,d\star\alpha$ (or $\square\alpha=d\star d\alpha)$.

\subsection{Maxwell equations}

One of the most beautiful application of the exterior form formalism
is that it gives a clean and unified picture of the classical theory
of electromagnetism \cite{flanders1963differential,frankel2011geometry}.
Here are the Maxwell equations written with differential forms where
we drop the constant $\epsilon_{0},\mu_{0},c$. Maxwell Theory is
given by the wonderful table 

\[
\boxed{\begin{array}{ccccccccc}
\Lambda^{0}(\mathbb{R}^{4}) & \xrightleftharpoons[\partial]{d} & \Lambda^{1}(\mathbb{R}^{4}) & \xrightleftharpoons[\partial]{d} & \Lambda^{2}(\mathbb{R}^{4}) & \xrightleftharpoons[\partial]{d} & \Lambda^{3}(\mathbb{R}^{4}) & \xrightleftharpoons[\partial]{d} & \Lambda^{4}(\mathbb{R}^{4})\\
\\
f & \xrightleftharpoons[(2)]{(1)} & (V,A) & \xrightarrow{(3)} & (E,B) & \xrightarrow{(4)} & 0\\
 &  &  &  & \star(E,B) & \xrightarrow[(5)]{} & (\rho,j) & \xrightarrow[(6)]{} & 0
\end{array}}
\]
Notice that the dimension of $\Lambda^{1}(\mathbb{R}^{4})$, $\Lambda^{2}(\mathbb{R}^{4})$
and $\Lambda^{3}(\mathbb{R}^{4})$ are $4$, $6$ and $4$. We denote
here
\begin{itemize}
\item $A^{\mu}=(V,A)$ : the potential and potential vector (usually a $4-$vector).
This is a $1-$form as it is linked to the phase of the wave function
of the electron (as for example in the Aharonov Bohm effect).
\item $F=(E,B)$ : the electromagnetic field (usually already a antisymmetric
$2-$tensor). This is a $2-$form, one would want to integrate the
magnetic field on a surface for example.
\item $J^{\mu}=(\rho,j)$ : charge density and current (usually a $4$-vector).
This is a $3-$form, the density is integrate on a volume and the
current on a surface times a time intervale. 
\end{itemize}
A great advantage here that everything is at the ``geometric'' level.
There is no choice of parametrisation of the space. No worry about
the change of referencial.

Here every arrow of the diagramm is just the exterior derivative.
Bellow are their usual meaning:
\begin{enumerate}
\item[(1)] This is a Gauge invariance, one can change 
\[
V\rightarrow V-\frac{\partial f}{\partial t}\quad\text{et}\quad A\rightarrow A+\text{grad}(f)
\]
without modifying electromagnetic field, indeed $(3)\circ(1)=0$. 
\item[(2)] A particular choice of Gauge called Lorentz's Gauge $(2)=0$ that
is 
\[
\frac{\partial V}{\partial t}+\text{div}(A)=0.
\]
\item[(3)] Electromagnetic fields are expressed with the potential and potential
vector: 
\[
E=-\frac{\partial A}{\partial t}-\text{grad}(V)\quad\text{et}\quad B=\text{curl}(A).
\]
\item[(4)] Here are the Maxwell-Faraday et Maxwell-Thomson equation
\[
\frac{\partial B}{\partial t}+\text{curl}(E)=0\quad\text{et}\quad\text{div}(B)=0
\]
This follows of course from $(4)\circ(3)=0$.
\item[(5)] Here are now Maxwell-Gauss and Maxwell-Ampere equations
\[
\text{div}(E)=\rho\quad\text{et}\quad-\frac{\partial E}{\partial t}+\text{curl}(B)=j
\]
\item[(6)]  This is the conservation law of the charge
\[
\frac{\partial\rho}{\partial t}+\text{div}(j)=0
\]
Again that is $(6)\circ(5)=0$.
\end{enumerate}
We can finish by a small historical remark, in 1865 the fantastic
idea of Maxwell were to notice that $(6)\circ(5)\neq0$ with Ampere
equation as stated at that time. He modified the equation adding the
term $\frac{\partial E}{\partial t}$ to obtain a coherent theory.
Therefore it is indeed a differential geometry approach that gave
nowday's classical electromagnetic theory. 

\section{De Rham (Trivial) Cohomologie }

We have seen in Section \ref{sec:Exterior-Derivative} that if a form
$\alpha$ can be written as $\alpha=d\beta$ it satisfies $d\alpha=0$.
A natural question would be to asked whether the converse is always
true. The answer is yes for $\mathbb{R}^{n}$.
\begin{prop}
\label{prop:(De-Rham-(trivial)}(De Rham (trivial) Cohomology) In
$\mathbb{R}^{n}$, for any form $\alpha$ that is not a constant $0-$form
\[
d\alpha=0\quad\Leftrightarrow\quad\text{there exists }\beta\text{ such that }\alpha=d\beta.
\]
\end{prop}

This proposition is not true for more complicated topological space.
It gives rize to a whole domain of study called De Rham Cohomology
which happens to be one of the most powerful tools to study and charaterize
topological object in differential geometry and algebraic topology\footnote{There, the formula $\phi_{t}^{*}(\alpha)=\alpha+dI_{t}\alpha$ is
still very useful since it implies that $\alpha$ and $\phi_{t}^{*}\alpha$
belongs to the same ``class''.} \cite{CodbillonElementTopologie}. But in $\mathbb{R}^{3}$ all this
is very well known by any undergrad student :
\begin{itemize}
\item $\alpha$ is a 0-form : $\text{grad}(\alpha)=0$ iff there exists
$c\in\mathbb{R}$ constant such that $\alpha=c$, 
\item $\alpha$ is a 1-form : $\text{curl}(\alpha)=0$ iff there exists
a $0-$form $\beta$ such that $\alpha=\text{grad}(\beta)$,
\item $\alpha$ is a 2-form : $\text{div}(\alpha)=0$ iff there exists a
$1-$form $\beta$ such that $\alpha=\text{curl}(\beta)$, 
\item $\alpha$ is a 3-form : The equation $\text{div}(\beta)=\alpha$ always
has a solution.
\end{itemize}
Notice that it is possible to construct $\beta$ explicitely. For
example, if $\alpha$ is a 1-form, define the $0$ form $f(x)=\int_{x_{0}}^{x}\alpha$
with $x_{0}$ fixed and then $\alpha=\text{grad}(f)$. Or if $\rho_{0}$
is a 3-form seen as the density of mass that has been transported
from far away, we have with the continuous equation
\[
\rho_{0}=-\int_{-\infty}^{0}\text{div}(\vec{v}_{t}\rho_{t})dt=\text{div}\left(-\int_{-\infty}^{0}\vec{v}_{t}\rho_{t}dt\right).
\]

\begin{figure}
\centering{}\includegraphics[width=4cm]{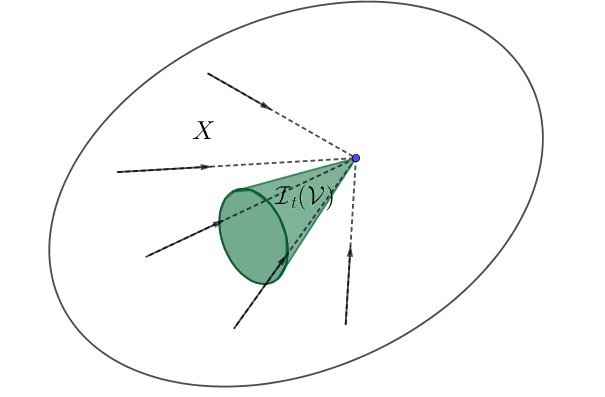}\includegraphics[width=4cm]{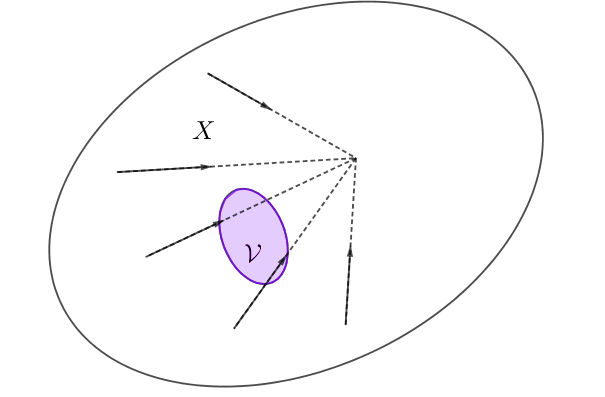}\includegraphics[width=4cm]{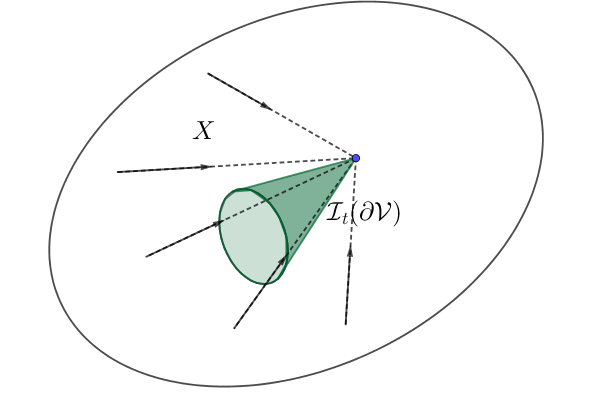}\caption{\label{enlarged-submanifold-1-1} A ``Geometric proof'' of $d\alpha=0$
$\Rightarrow$ $\alpha=d\beta$. Choose a tangent vector field which
induces a contraction of the space to a single point. Then in Cartan's
magic formula, the terms $\alpha({\cal V}(t))$ and $\alpha(\partial(I_{t}{\cal V}))$
vanish and we are left with $0=-\alpha({\cal V})-\alpha({\cal I}_{t}(\partial{\cal V}))$.
Therefore $\alpha=d(-I_{t}\alpha).$}
\end{figure}

In higher dimension, It also implies important properties.
\begin{cor}
If $J$ is a conserved quantity, then there exists $\beta$ such that
$d\beta=J$.\label{cor:associated_field}
\end{cor}

In particular in $\mathbb{R}\times\mathbb{R}^{3}$ for any $(\rho,j)$
that satisfies the continuous equation there exists a field $\beta=({\cal E},{\cal B})$
for which Maxwell-Gauss and Maxwell-Ampere equations are valid
\[
\text{div}({\cal E})=\rho\quad\text{and}\quad-\frac{\partial{\cal E}}{\partial t}+\text{curl}({\cal B})=j.
\]
This is a purely mathematical construction. This field should be seen
as an equivalent to the potential flow that describes an irrotational
velocity field as a gradien in hydrodynamics.

Also such a $\beta$ is not unique and one can made a choice of Gauge.
In the case $d(\star\beta)=0$ we have the following 
\begin{cor}
\label{cor:wave_equation}If $J$ is a conserved quantity, then there
exists $\alpha$ such that 
\[
d\star d\alpha=J.
\]
\end{cor}

In a coordinate system it is the usual propagation equation with a
source

\begin{equation}
\square\alpha=J\label{eq:propagation}
\end{equation}
We should stress that this is true for \emph{any} conserved quantity
with no more information about the physics of the system. However
there are important examples where $\alpha$ appears in theorical
model or is indeed a real physical quantity. For example :
\begin{itemize}
\item The electric charge is conserved. We have the electromagnetic field
: (\ref{eq:propagation}) with $J=(\rho,j)$ the charge density and
current and $\alpha=(V,A)$ potential and vector potential.
\end{itemize}
So we can ask the following.
\begin{question}
\label{que:associatedField}What is the associated $\alpha$ for the
conservation of energy, momentum, moment of inertia, weak charge,...
? 
\end{question}

\section{Euler Lagrange Equations}

This last section is more ``theorical physics'' than ``daily physics''
as we will focus on the Lagrangian approach for a classical field
theory and its very nice consequences such as Noether Theorem. Again
we will express everything using exterior forms.

\subsection{Euler Lagrange Equations}

As a most simple model the Lagrangian ${\cal L}$ is a $n$-form\footnote{Most commun conventions call the Lagrangian $\star{\cal L}$ which
is then a $0$-form (a scalar).} and we assume the following.
\begin{assumption}
\label{assu:form-Lagrangian}${\cal L}={\cal L}(\alpha,d\alpha)$
only depends only on a $k-$form $\alpha$ and its exterior derivative
$d\alpha$. 
\end{assumption}

For $k\geq1$, the last hypothesis is more restrictive than considering
all the derivatives $(\frac{\partial\alpha}{\partial x_{i}})_{1\leq i\leq n}$
but is also reasonable if we believe that $d\alpha$ has more geometric
or physical meaning\footnote{If $\alpha$ is a $0$-form then considering $d\alpha$ is similar
to considering all the derivatives $(\frac{\partial\alpha}{\partial x_{i}})_{1\leq i\leq n}$.}. Also a nice aspect of this formalism is that it gives a justification
why the second order derivatives $\frac{\partial^{2}\alpha}{\partial x_{i}\partial x_{j}}$
are not used in the Lagrangian : indeed $d\circ d\alpha=0$.

We write ${\cal L}_{\alpha}:=\partial_{1}{\cal L}(\alpha,d\alpha)$
that is an $(n-k)$-form and ${\cal L}_{d\alpha}:=\partial_{2}{\cal L}(\alpha,d\alpha)$
that is a $(n-k-1)$-form such that for a small perturbation $\delta\alpha$
at first order we have
\begin{align*}
{\cal L}(\alpha+\delta\alpha,d\alpha+d\delta\alpha) & \approx{\cal L}(\alpha,d\alpha)+\delta\alpha\wedge{\cal L}_{\alpha}+(d\delta\alpha)\wedge{\cal L}_{d\alpha}
\end{align*}
With this formulation we also assume that ${\cal L}$ depends only
locally on $\alpha$, in the sense that if $\delta\alpha$ is supported
on a region $\Omega'\subset\mathbb{R}^{n}$ then ${\cal L}$ is not
modified outside of $\Omega'$. We now write down the Euler-Lagrange
Equation. Remark that we have
\begin{align*}
{\cal L}(\alpha+\delta\alpha,d\alpha+d\delta\alpha) & \approx{\cal L}(\alpha,d\alpha)+\delta\alpha\wedge({\cal L}_{\alpha}-(-1)^{k}d{\cal L}_{d\alpha})+d(\delta\alpha\wedge{\cal L}_{d\alpha}).
\end{align*}
For a submanifold ${\cal V}$, in order to maximise $\int_{{\cal V}}{\cal L}(\alpha,d\alpha)$
with fixed boundary conditions on $\partial{\cal V}$ for $\alpha$
we obtain the following.
\begin{defn}
(Euler Lagrange Equation) 
\[
{\cal L}_{\alpha}-(-1)^{k}d{\cal L}_{d\alpha}=0
\]
\end{defn}

For $\alpha$ a $0$-form this is the usual Euler Lagrange Equation
in Field Theory 
\[
\frac{\partial{\cal L}}{\partial\alpha}=\partial_{\mu}\left(\frac{\partial{\cal L}}{\partial(\partial_{\mu}\alpha)}\right).
\]
A nice example with $\alpha$ a $1$-form is of course the electromagnetic
field with $\alpha=(V,A)$ and $J=(\rho,j)$ we have
\begin{defn}
(Electromagnetism Lagrangian)
\[
{\cal L}^{(\text{EM})}=\alpha\wedge J-\frac{1}{2}(d\alpha\wedge\star d\alpha).
\]
Here ${\cal L}_{\alpha}^{(\text{EM})}=J$, ${\cal L}_{d\alpha}^{(\text{EM})}=\star d\alpha$
and the Euler Lagrange Equations gives again the Maxwell equations
: 
\[
d\star d\alpha=J
\]
\end{defn}

\subsection{A conserved quantity in the $1$-Form case}

Remark that Euler Lagrange Equation directly implies that $d{\cal L}_{\alpha}=0$.
So in particular we have the following. 
\begin{cor}
\label{cor:conserve-1f}If $\alpha$ is a 1-form then ${\cal L}_{\alpha}$
is a conserved quantity and ${\cal L}_{d\alpha}$ is an associated
field\footnote{As in Corollary \ref{cor:associated_field}}.
\end{cor}

One way to understand Corollary \ref{cor:conserve-1f} is that assumption
\ref{assu:form-Lagrangian} add a Gauge symmetry to the system. For
example with a perturbation of the form $\delta\alpha=d(\delta f)$
we have $d(\delta\alpha)=0$ and then 
\begin{align*}
{\cal L}(\alpha+d\delta f,d\alpha) & \approx{\cal L}(\alpha,d\alpha)-\delta f\wedge d{\cal L}_{\alpha}+d(\delta f\wedge{\cal L}_{\alpha}).
\end{align*}
and finally obtain $d{\cal L}_{\alpha}=0$ as for the Euler Lagrange
Equation. Again we have a nice example with the electromagnetic field
where ${\cal L}_{\alpha}^{(\text{EM})}=J$ is a conserved quantity
and $d{\cal L}_{d\alpha}^{(\text{EM})}=J$.

\subsection{Noether Theorem}

We now mention the famous Noether Theorem. Remark that Euler Lagrange
equation implies that 
\begin{align*}
{\cal L}(\alpha+\delta\alpha,d\alpha+d\delta\alpha) & -{\cal L}(\alpha,d\alpha)\approx d(\delta\alpha\wedge{\cal L}_{d\alpha})
\end{align*}
We denote $\xi$ an infinitesimal transformation $\alpha\rightarrow\alpha+\delta\alpha^{\xi}$
and ${\cal L}\rightarrow{\cal L}+\delta{\cal L}^{\xi}$. If the system
is invariant by this transformation $\delta{\cal L}^{\xi}=0$ then
$d(\delta\alpha^{\xi}\wedge{\cal L}_{d\alpha})=0$ so $\delta\alpha^{\xi}\wedge{\cal L}_{d\alpha}$
is a conserved quantity. We can state a more general result \cite{olver1993applications}.
\begin{thm}
(Noether Theorem) If $\delta{\cal L}^{\xi}=d(\delta\Lambda^{\xi})$
then $\delta\alpha^{\xi}\wedge{\cal L}_{d\alpha}-\delta\Lambda^{\xi}$
is a conserved quantity.
\end{thm}

A very nice application of Noether Theorem is of course the conservation
of energy and momentum.

\subsection{The stress-energy tensor}

We consider translations of the system and more generaly the transport
along a flow given by vector field $X$. Notice that we have $L_{X}{\cal L}=d(i_{X}{\cal L})$
(Cartan's magic formula) so we can apply Noether Theorem. We compute
\[
L_{X}{\cal L}=L_{X}\alpha\wedge{\cal L}_{\alpha}+(dL_{X}\alpha)\wedge{\cal L}_{d\alpha}+L_{X}{\cal L}|_{\alpha,d\alpha}
\]
where the last term is the derivative of ${\cal L}$ assuming $\alpha,d\alpha$
fixed and then we obtain
\begin{align*}
d(L_{X}\alpha\wedge{\cal L}_{d\alpha}-i_{X}{\cal L}) & =-L_{X}{\cal L}|_{\alpha,d\alpha}.
\end{align*}
For example in the case of the electromagnetism Lagrangian and $X=\partial_{t}$
for translation in time we obtain Poynting's theorem
\[
\partial_{t}(\frac{1}{2}(|E|^{2}+|B|^{2})+\vec{A}.\vec{j})+\text{div}(E\times B)=\vec{A}.\partial_{t}\vec{j}.
\]
For more general Lagrangian ${\cal L}$ we can define the following.
\begin{defn}
We call $L_{X}\alpha\wedge{\cal L}_{d\alpha}-i_{X}{\cal L}$ the \emph{stress-energy
tensor} if $X$ is a translation.
\end{defn}

In a coordinate system and with the translations $X=\partial_{\nu}$,
this is the usual formula for the stress-energy tensor defined from
a Lagrangian
\[
T_{\mu\nu}=\partial_{\nu}\alpha^{\eta}\left(\frac{\partial{\cal L}}{\partial(\partial_{\mu}\alpha)}\right)_{\eta}-\eta_{\mu\nu}{\cal L}.
\]
And we also state the conservation of energy and momentum in a general
setting.
\begin{cor}
If $L_{X}{\cal L}|_{\alpha,d\alpha}=0$, ie. the Lagrangian is invariant
by the transformation induced by the vector field $X$, then $(L_{X}\alpha)\wedge{\cal L}_{d\alpha}-i_{X}{\cal L}$
is a conserved quantity. 
\end{cor}

\subsection{Hamilton Equation}

The energy tensor 
\[
{\cal H}=L_{X}\alpha\wedge{\cal L}_{d\alpha}-i_{X}{\cal L}.
\]
corresponds to the definition of the Hamiltonian in the case $X=\partial_{t}$.
With some computation we obtain for small perturbation
\begin{align*}
\delta{\cal H} & =L_{X}\alpha\wedge\delta{\cal L}_{d\alpha}-\delta\alpha\wedge L_{X}{\cal L}_{d\alpha}-di_{X}(\delta\alpha\wedge{\cal L}_{d\alpha})\\
 & -i_{X}(\delta\alpha\wedge({\cal L}_{\alpha}-(-1)^{k}d{\cal L}_{d\alpha}))
\end{align*}
There if we assume that ${\cal H}={\cal H}(\alpha,{\cal L}_{d\alpha})$
and that the small perturbation is given by 
\[
\delta{\cal H}=\delta\alpha\wedge H_{1}+\delta{\cal L}_{d\alpha}\wedge H_{2}
\]
we obtain the Hamilton Equations.
\begin{prop}
(Hamilton Equations) Euler Lagrange Equations are equivalent to
\[
H_{1}=-L_{X}{\cal L}_{d\alpha}\quad\text{and}\quad H_{2}=(-1)^{k+1}L_{X}\alpha.
\]
\end{prop}

\begin{rem}
The stationary form (for which $L_{X}\alpha=0$ and then $L_{X}{\cal L}_{2}=0$)
which that are critical point of the Hamiltonian ($\delta H=0$ for
any $\delta\alpha$) solve Euler Lagrange equation. In particular
if $\alpha$ is such that it minimizes the energy. 
\end{rem}

\subsection{Gravitational waves (?)}

We finish by giving a partial answer for Question \ref{que:associatedField}.
One can find in a book on General Relativity this equation\footnote{Only true in the weak gravitation regime where Einstein equation can
be linearized. } used to describe gravitational waves 
\[
\square\tilde{h}=T
\]
where $\tilde{h}$ is constructed with the perturbation of the metric
$g$ around the flat Minkowski metric and $T$ is the $4\times4$
Stress-Energy tensor. Here one can think of each line of $T$ as a
3-form which corresponds to the conservation of energy (first line)
and the conservation of momentum (the three others lines) and then
the line of $\tilde{h}$ play the role of the associated $\alpha$
in Corollary \ref{cor:wave_equation} and the associated $\beta$
in Corollary \ref{cor:associated_field} can be interpreted as the
classical gravitational field.

Unfortunatly general relativity is much more complicated and the above
statement is true only at first order.

\section{Summary table}
\begin{center}
\begin{tabular}{|c|c|}
\hline 
$m^{-2}s^{-1}$ & $dx\wedge dy\wedge dt$\tabularnewline
\hline 
gradient, curl, divergence & $d$\tabularnewline
\hline 
irrotational, divengence free, conserved quantity & $d\alpha=0$\tabularnewline
\hline 
Gauge invariance & $d(\alpha+d\beta)=d\alpha$\tabularnewline
\hline 
De Rham cohomology & $d\alpha=0\Rightarrow\alpha=d\beta$ ?\tabularnewline
\hline 
Hydrodynamics & $\partial_{t}\alpha+L_{V}\alpha=...$\tabularnewline
\hline 
Maxwell equations & $dF=0,$  $\quad d\star F=0$\tabularnewline
\hline 
Propagation wave with source & $d\star d\alpha=J$\tabularnewline
\hline 
Euler Lagrange equation & ${\cal L}_{\alpha}-(-1)^{k}d{\cal L}_{d\alpha}=0$\tabularnewline
\hline 
Stress-energy tensor & $L_{X}\alpha\wedge{\cal L}_{d\alpha}-i_{X}{\cal L}$\tabularnewline
\hline 
\end{tabular}
\par\end{center}

\bibliographystyle{alpha}
\bibliography{../../Documents/Travail_de_recherche/Other_small_projects/biblio_Forme_Diff}

\end{document}